\documentclass[%
 preprint,
 amsmath,amssymb,
 aps,
pra,
]{revtex4-2}

\usepackage{etoolbox} 

\usepackage{graphicx}
\usepackage{xcolor}
\usepackage{bm}

\begin{document}

\title{Quantum Thermodynamics on a limit cycle. }

\author{Varinder Singh}
\affiliation{School of Physics, Korea Institute for Advanced Study, Seoul 02455, Republic of  Korea}

\author{Euijoon Kwon}
\affiliation{Department of Physics and Astronomy \& Center for Theoretical Physics, Seoul National University, Seoul 08826, Republic of Korea} 
\affiliation{School of Physics, Korea Institute for Advanced Study, Seoul 02455, Republic of  Korea}

\author{G J Milburn}
 \affiliation{Sussex Centre for Quantum Technologies, University of Sussex, Brighton, BN1 9RH, UK}

\date{\today}

\begin{abstract}
We consider a periodic quantum clock based on cooperative resonance fluorescence at zero temperature. The semi-classical dynamics exhibit a Hopf bifurcation to a limit cycle. In the quantum case, this system has an exact steady state and the limit cycle appears in conditional quantum dynamics under homodyne detection. We show that the intrinsic quantum phase diffusion on the limit cycle leads to fluctuations in the period. By simulating the stochastic master equation for homodyne detection, we extract the statistical properties of the clock period. We show that the precision of the clock satisfies the quantum-thermodynamic kinetic uncertainty relations. As energy dissipation increases, the clock’s quality improves, fully validating in a quantum stochastic system the link between energy dissipation and clock precision.   
\end{abstract}

\maketitle


\section{Introduction}
\label{intro}
In classical mechanics, limit cycles arise in a driven, and dissipative, nonlinear system when a fixed point undergoes a Hopf bifurcation leading to a one dimensional attractor\cite{limit-cycles}. This forms when the average work done on the system over the limit cycle equals the average energy dissipated. However, the fluctuation-dissipation theorem requires that noise accompanies dissipation. In the case of a limit cycle, this noise induces phase diffusion, leading to fluctuations in the oscillation period \cite{Aminzare}.  

Recently W\"{a}chtler et al. studied noise and fluctuations in the single electron shuttle \cite{Waechtler}. The single electron shuttle is a nano-mechanical system in which electron tunnelling though a quantum dot is coupled to the mechanical motion of a resonator in which the dot resides \cite{Gorelik}. It exhibits a rich nonlinear dynamical structure, including a Hopf bifurcation that leads to a limit cycle. \cite{Dian}.  The analysis was carried out for the case in which thermal fluctuations dominate the tunnelling, the typical experimental situation\cite{Ares}. However quantum tunnelling can take place at `zero' temperature (viz.  in which thermal fluctuations are negligible compared to dissipative quantum tunnelling) and thus it is of interest to investigate the thermodynamics fluctuation theorems in this limit. In this paper we study such an example drawn from quantum optics in which the zero temperature approximation is very good. 

The occurrence of a Hopf bifurcation in a dynamical system indicates a possible candidate for a clock if the oscillations can be reliably counted. Our analysis shows that the model of collective resonance fluorescence is a good example of a how a  quantum dissipative system with a semiclassical Hopf bifurcation may be used as a quantum clock.  Our analysis uses quantum kinetic uncertainty relations, one of the key results in the field of quantum stochastic thermodynamics. These relations show that quantum coherence can improve clock performance\cite{Manikandan:2022mfk,Stefan1, PhysRevResearch.7.013077}. In the model of this paper, measurement back action drives the system away from its steady state and creates the quantum coherence needed for periodic dynamics.

\section{Collective resonance fluorescence}
The Dicke superradiance model \cite{Dicke1954} is a well understood phenomenon in quantum optics. A collection of  $N$  two-level atoms is confined to a small region of space, ensuring that they all experience approximately the same laser field . This results in a collective spontaneous emission enhanced by the factor $N$. We will follow the treatment of Drummond and Hassan \cite{DrummondHassan1980}. 

Each atomic system is described by a two level transition between a ground state, $|g\rangle$ and an excited state, $|e\rangle$. The full Hilbert space is $2^N$ dimensional but in the Dicke model we are confined to the symmetric subspace of dimension $N+1E$. In that case we can introduce the collective $su(2)$ operators as 
\begin{equation}
\hat{J}_\alpha =\sum_{i=1}^N \hat{j}_\alpha^i
\end{equation}
where $\alpha=\{x,y,z\}$ and we define the single atom spin operators 
\begin{eqnarray}
\hat{j}_x & = & \frac{1}{2}(|e\rangle\langle g|+|g\rangle\langle e|)\\
\hat{j}_y & = & -\frac{i}{2}(|e\rangle\langle g|-|g\rangle\langle e|)\\
\hat{j}_z & = & \frac{1}{2}(|e\rangle\langle e|-|g\rangle\langle g|)
\end{eqnarray}
so that we have the usual $su(2)$ algebra, $[\hat{J}_x, \hat{J}_y]=i\hat{J}_z$.  The total system obeys the master equation, in the interaction picture, 
\begin{equation}
\label{SR-me}
\dot{\rho}= -i\Omega[\hat{J}_x,\rho]+\gamma{\cal D}[\hat{J}_-]\rho={\cal L}\rho
\end{equation}
where $\hat{J}_\pm = \hat{J}_x\pm i \hat{J}_y$ and $\Omega$ and $\gamma$ are the Rabi frequency and spontaneous emission rate respectively. We have assumed that the atomic transition is driven on resonance by an sinusoidal coherent field that does work on the system. The damping is due entirely to spontaneous emission into a zero temperature heat bath; a very good approximation at optical frequencies.

The master equation implies that the first order moments are coupled to higher order moments,
\begin{eqnarray}
\frac{d\langle \hat{J}_+\rangle}{dt} & = & -i\Omega\langle \hat{J}_z\rangle
+\gamma\langle \hat{J}_+ \hat{J}_z\rangle\\
\frac{d\langle \hat{J}_z\rangle}{dt} & = & \Omega\langle \hat{J}_y\rangle -\gamma\langle \hat{J}_+ \hat{J}_-\rangle
\end{eqnarray}
 Following Carmichael\cite{Carmichael}, we define the scaled moments 
\begin{equation}
X+iY = \langle \hat{J}_-\rangle/N ,\ \ \  Z=\langle \hat{J}_z\rangle/N
\end{equation} 

The semiclassical equations of motion for the first order moments are given by 
\begin{eqnarray}
dX & = & \gamma N Z X dt\label{semi-classical1}\\
dY & = & -\Omega Z dt +\gamma NY Z dt\label{semi-classical2}\\
dZ & = & \Omega Y dt-\gamma N (1/4-Z^2)dt-\frac{\gamma}{4}(1+2 Z)^2dt\label{semi-classical3}
\end{eqnarray}
where we have factorised second order moments (i.e. neglecting quantum correlations).  The semiclassical equaitons result when we let $\gamma\rightarrow 0, N\rightarrow \infty$ such that $N\gamma=\mbox{constant}$. In this limit we can neglect the last term in Eq. (\ref{semi-classical3}). We will assume this for the rest of this section.

It is easy to confirm that $|\langle J_+\rangle|^2+\langle J_z\rangle^2$ is a constant of the motion (neglecting the last term in Eq. (\ref{semi-classical3}))  and the constant is $N^2/4$.  We now use a stereo-graphic projection of the Bloch sphere defined by 
\begin{eqnarray}
\langle \hat{J}_+\rangle & = & \frac{2Jz(t)}{1+|z|^2}\\
 \langle \hat{J}_z\rangle & = &  J\frac{(1-|z|^2)}{(1+|z|^2)}
\end{eqnarray}
In this projection the excited state is projected to the origin, the ground state to infinity and the equator to the unit circle.   
The semiclassical equations of motion can be written as 
\begin{equation}
\frac{dz}{dt} = i\Omega(z^2-1)/2+\gamma Jz
\end{equation}
where $J=N/2$.  The general solution to this equation is given in \cite{DrummondHassan1980}. 
 As $\Omega$ increases from zero the stable fixed point starting at $\langle J_z\rangle =-N/2$ is rotated around the $x$ axis until it reaches the equatorial plane.   For $\Omega >\Omega_0=\gamma N/2$ there is a Hopf bifurcation giving rise to a stable limit cycle with frequency $\omega= \sqrt{\Omega^2-\Omega_0^2}$.   In this paper we will always start with $\langle \hat{J}_x(t)\rangle =0$. In that case the motion in the stereo-graphic plane is confined to $z(t)=iu(t)$ with $u(t)$ real. On the limit cycle the solution is 
\begin{equation}
    u(t)=\Im \left [ \frac{\mu+\mu^*e^{i\omega t/2}} {1-\cos(\omega t/2)}\right ]
\end{equation}
\begin{equation}
 \mu =\left [1- \frac{\Omega_0^2}{\Omega^2}\right ]^{1/2}+i\frac{\Omega_0}{\Omega}   
\end{equation}
where the frequency on the limit cycle is $\omega =(\Omega^2-\Omega_0^2)^{1/2}$.
The limit cycle is confined to a circle in the $y-z$ plane with the radius given by $N/2$. 

This suggests we make a change of variable so that
\begin{eqnarray}
\langle \hat{J}_x(t)\rangle & = & 0 \label{semiclassical-solutions1}\\
\langle \hat{J}_y(t)\rangle & = & \frac{N}{2} \sin(\theta(t))\label{semiclassical-solutions2}\\
\langle \hat{J}_z(t)\rangle & = & \frac{N}{2} \cos(\theta(t))\label{semiclassical-solutions3}
\end{eqnarray}
Substituting this into the semiclassical equations we find that
\begin{equation}
   \dot{\theta}= -\Omega +\Omega_0\sin{\theta} 
\end{equation}
The solution is
\begin{equation}
 \theta(t)=2\arctan\left [a-b\tan[\omega (t-K)/2]\right ]   
\end{equation}
where $K$ is constant determined by the initial condition. We will take $\theta(0)=\pi/2$ in which case 
\begin{equation}
  K=\frac{2}{\Omega b}\arctan((1-a)/b)  
\end{equation}
where $a=\Omega_0/\Omega$ and $b=\sqrt{1-a^2}$.

\begin{figure}[h!]
\centering
\includegraphics[scale=0.5]{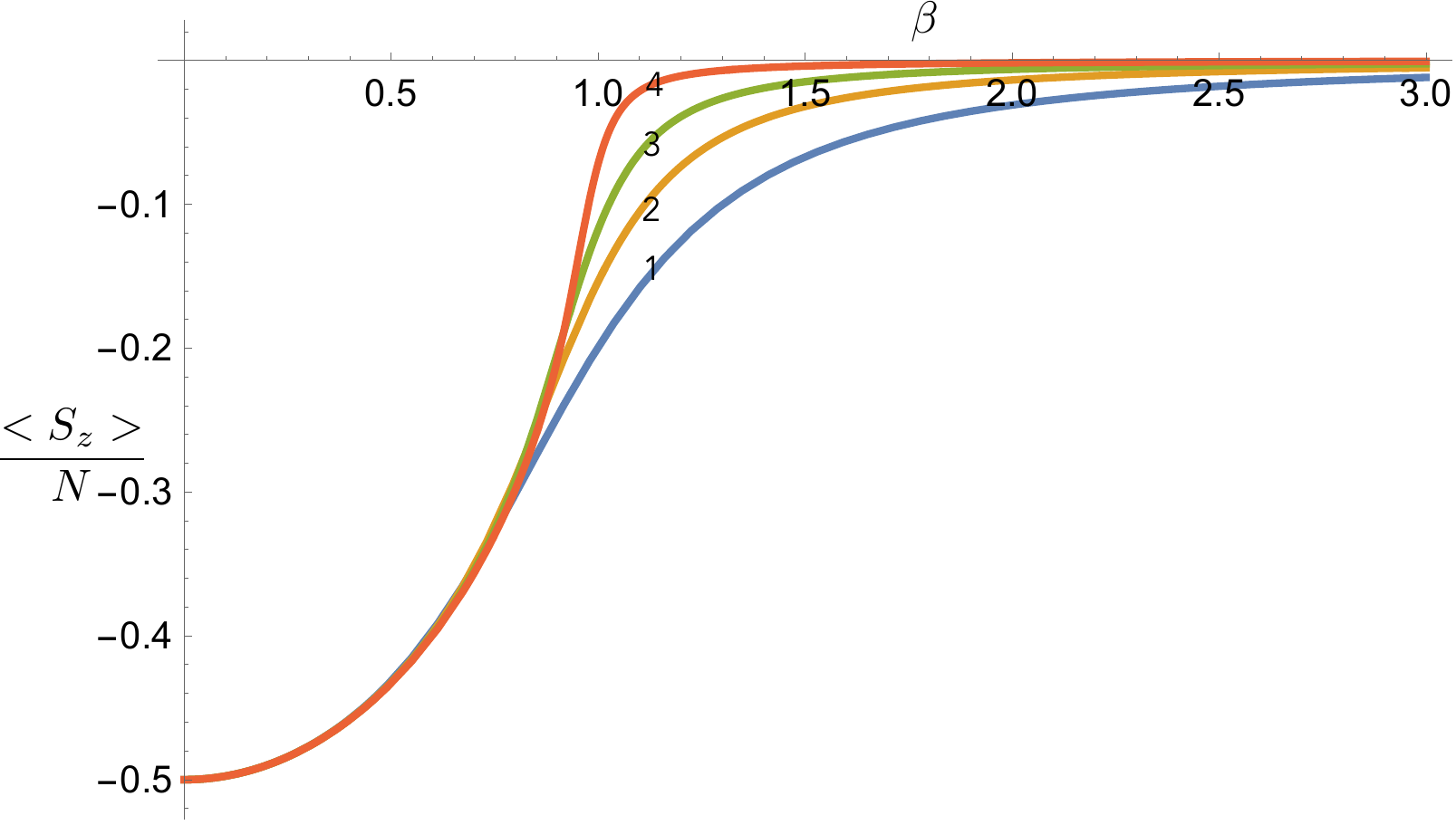}
\caption{The exact steady state value for $\langle \hat{J}_z\rangle/N$ as a function of $\beta=\Omega/\Omega_0$. The four cases correspond to (1) $N=5$, (2) $N=10$, (3) $N=20$, (4) $N=70$.}
\label{PL-fig1}
\end{figure}

The exact quantum steady state solution to the master equation is known\cite{Puri},
\begin{equation}
\rho_{ss} = {\cal N}\sum_{m,n=0}^{2J} (i\Omega/\gamma)^{2J-m}(-i\Omega/\gamma)^{2J-n}(\hat{J}_-)^m(\hat{J}_+)^n
\end{equation}
where ${\cal N}$ is a normalisation constant. This may be used to compute the exact steady state moments as a function of $\Omega/\gamma$. In Fig.(\ref{PL-fig1}) we show the steady state value of $\langle \hat{J}_z\rangle/N$ as a function of $\beta=\Omega/\Omega_0$. 
The curve is continuous and $\langle \hat{J}_z\rangle/N$ tends to zero above the Hopf bifurcation. How do we reconcile this with the semiclassical limit cycle for $\Omega >\Omega_0$?  

The reconciliation is apparent  when we include the quantum noise which causes phase diffusion on the limit cycle above threshold. The steady state is an ensemble average over all time and this averages the phase diffusion to zero around the limit cycle. The steady state is thus completely phase diffused around the equator of the Bloch sphere. While at first sight it appears that the steady state of $\langle \hat{J}_z\rangle/N$ as a function of $\beta$ in the limit of large $N$ would fill the role of a threshold function, the dynamics is in fact more complicated in that there is a limit cycle not a fixed point above the threshold. How does this manifest in a physical realisation of the system?

To answer this we need to consider how we measure this system to see the threshold behaviour. We will use the theory of weak continuous measurements. The average field radiated by the collective dipole is \begin{equation}
\hat{E}_{out}(t) =\sqrt{\gamma} \hat{J}_-(t)-\hat{E}_{in}(t)
\end{equation}
where the input field  is in the vacuum state
\begin{equation}
\langle \hat{E}_{in}(t) \hat{E}_{in}^\dagger(t')\rangle =\delta(t-t')
\end{equation}
We will assume that all the fluorescent light is collected and input to a homodyne detection scheme\cite{WM}.
The measured signal is a homodyne current that is conditioned on the state of the source. It is given by the Ito stochastic differential equation,
\begin{equation}
J(t)dt =\gamma\langle J_-(t)e^{i\phi}+J_+(t)e^{-i\phi}\rangle_c dt +\sqrt{\gamma}dW(t) 
\end{equation}
where $dW(t)$ is the Wiener process and $\phi$ is the phase of the local oscillator of the homodyne detector. The quantum average in this expression is a conditional averages using  the conditional density operator of the system conditioned on the past measurement record to time $t$. This is given by a stochastic master equation,
\begin{equation} \label{eq:SME}
d\rho_c ={\cal L}\rho_c dt +\sqrt{\gamma}dW(t){\cal H}[\hat{J}_-e^{i\phi}]\rho_c
\end{equation}
where ${\cal L}$ is deined in Eq. (\ref{SR-me}) and the non linear conditionig operator is defined by 
\begin{equation}
{\cal H}[\hat{A}]\rho  = \hat{A}\rho+\rho \hat{A}^\dagger-{\rm tr}(\hat{A}\rho+\rho \hat{A}^\dagger)\rho
\end{equation} 

In the Bloch representation the stochastic master equation becomes
\begin{eqnarray*}
d\langle J_-\rangle_c & = & -i\Omega\langle J_z\rangle_cdt+\gamma\langle J_zJ_-\rangle_c dt\\
& &  +\sqrt{\gamma}(\langle J_-^2+ J_+J_-\rangle_c -\langle J_-\rangle_c(\langle J_++J_-\rangle_c) dW(t)\\
d\langle J_z\rangle_c &= & \Omega\langle J_y\rangle dt-\gamma\langle J_+ J_-\rangle_c dt\\
& & +\sqrt{\gamma}(\langle J_z J_-+J_+J_z\rangle_c -\langle J_z\rangle_c(\langle J_++J_-\rangle_c) dW(t)
\end{eqnarray*}

While the orbits are noisy, using the Ito calculus,  we see that they remain confined to the sphere defined by 
\begin{equation}
\langle \hat{J}_x^2+\hat{J}_y^2+\hat{J}_z^2\rangle_c =j(j+1)
\end{equation}
which indicates that the conditional state is a pure state. This is because the system starts in a pure state and no information is lost due to the measurement: every photon that is emitted is counted and no thermal photons enter the cavity.    The measurement creates the coherence necessary to see the limit cycle.  In an experiment of course there is loss of information as no photo detector is perfect. The conditional limit cycle is a closed curve on the Bloch sphere but due to fluctuations in $\hat{J}_x$  it is not confined to a circle in $y-z$ plane as in the semiclassical case.

\begin{figure}
\centering
\includegraphics[scale=0.5]{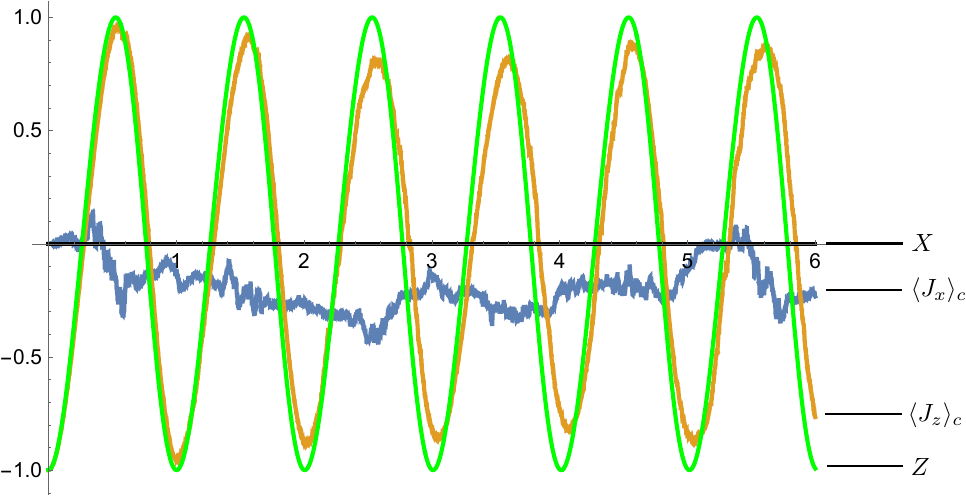}
\caption{A plot of the conditional dynamics of $\langle J_z(t)\rangle_c,\langle J_x(t)\rangle_c$ versus time using  homodyne detection of the fluorescent radiation with $N=10, \gamma=0.1,\Omega=2\pi$. Also shown is the semi-classical predictions $X,Z$. The phase diffusion in the conditional quantum dynamics is clear.   }
\label{quantum-condt}
\end{figure}

In Fig. (\ref{quantum-condt}) we plot the conditional mean values of $\langle \hat{J}_x\rangle_c, \langle \hat{J}_z\rangle_c$ as a  function of time conditioned on homodyne detection of the fluorescent field. The signature of the semiclassical limit cycle is apparent but with added amplitude and phase noise due to quantum measurement noise not thermal noise. As spontaneous emission makes the measurement possible we can regard the noise as due to spontaneous emission.  The quantum dynamics of $\langle \hat{J}_x\rangle_c$ ( see Fig.(\ref{quantum-condt})) differs considerably from the semiclassical dynamics of $X$ from Eq.(\ref{semi-classical3}) ; it appears as diffusive random process at a much slower time scale compared to the limit cycle oscillations.   

\begin{figure}
    \centering
    \includegraphics[scale=0.75]{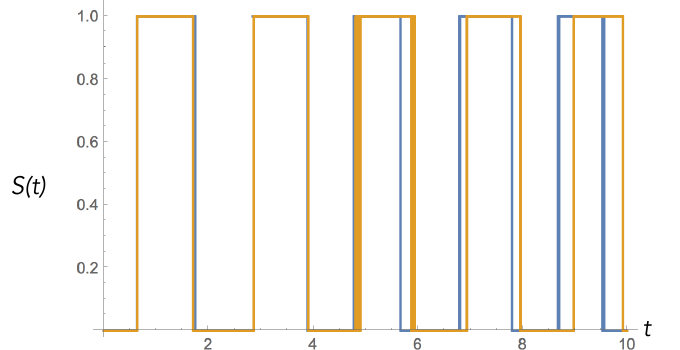}
    \caption{Two samples of the clock signal $s(t)$ with $j=12, \Omega=\pi, \gamma=1/j$}
    \label{clock-signal}
\end{figure}

In order to make apparent the fluctuations in the period, we define a clock `clock' signal as
\begin{equation}
    s(t)=(\mbox{sign}(\langle J_z(t)\rangle_c)+1)/2
\end{equation}
In Fig.(\ref{clock-signal}) we show two sample trajectories for clock signal.
The period is defined as the time between two consecutive falling edges in this record. Clearly this is a random variable. What is the probability distribution?

\section{Quantum Fluctuations and dissipation.}
The fluctuations of the period are determined the master equation, and  are measured by making phase dependent measurements on the system. Unlike the classical fluctuation theorems, the quantum fluctuation theorems require that we specify the continuously measured quantity in order to derive the conditional stochastic dynamics of the system. In the example of this paper, this measurement is homodyne detection of the fluorescent field emitted by a collection of $N$ two-level systems in the superradiance regime. 

We could also measure the intensity of the emitted light by making direct photon counting statistics on the output. The average count rate is proportional to $\langle J_+J_-\rangle$. In such a description we can derive a conditional dynamics, conditioned on the stochastic count in fixed time intervals.  

The quantum thermodynamic uncertainty relations can be used to quantify the quality of the clock based on homodyne detection.  In terms of quantum parameter estimation, the objective of a clock is to estimate the period of a system by integrating a continuous measurement record, in this case the homodyne current. This approach to parameter estimation was introduced by Gammelmark and Molmer \cite{PhysRevLett.112.170401}.  In the context of period estimation, this problem translates into a First Passage Time (FPT) problem. \cite{kewming2023passage}.  The results can be formulated in terms of a quantum kinetic  uncertainty relation \cite{VuSaito2022,Hasegawa2020}. 

\begin{figure}
        \centering
        \includegraphics[width=0.5\linewidth]{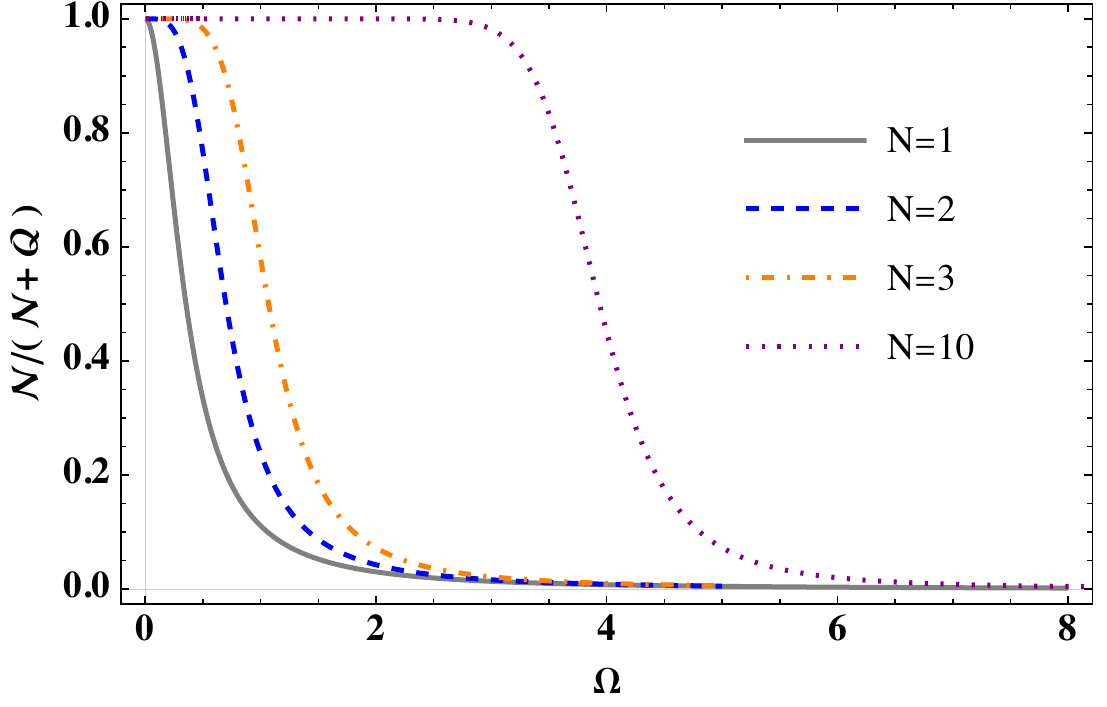}
        \caption{ $\mathcal{N}/(\mathcal{N}+\mathcal{Q})$ in Fig. (\ref{KURratio}) as a function of driving strength $\Omega$. Here, we have fixed $\gamma=1$. Solid gray, dashed orange, dot-dashed green and dotted red curves represent the case $N=1$, $N=2$, $N=3$ and $N=10$, respectively.}
        \label{KURratio}
    \end{figure}

For the master equation in Eq. (\ref{SR-me}), the FPT inequality takes the form
\begin{equation}
     \frac{{\rm Var}[T]}{\left(\partial_{\theta}{\rm E}_{\theta}[T]\vert_{\theta=0}\right)^{2}}\geq \frac{1}{I_{Q}(0)}\,.   
\end{equation}
where,  $\theta$ is the estimated parameter, and ${I_{Q}(0)=I_{Q}(\theta=0)}$ is the quantum Fisher information (QFI) \cite{Gammelmark_2014}. In the case considered here we get 
\begin{equation}
     I_{Q}(0) = \mathrm{E}[T](\overbrace{\mathcal{N}}^{classical} + \overbrace{\mathcal{Q}}^{quantum}), 
\end{equation}
where  $\mathcal{N}  =    Tr[J_+ J_- \rho_{\rm ss}] = \langle J_+ J_- \rangle,$ is the dynamical activity and $\mathcal{Q}$ is contribution from quantum coherence \cite{VuSaito2022}:
\begin{equation}
   \mathcal{Q} = -4 Tr [\mathcal{L}_1\ \mathcal{L}^\dagger\mathcal{L}_2\rho_{\rm ss}] -4 Tr[ \mathcal{L}_2\ \mathcal{L}^\dagger\mathcal{L}_1 \rho_{ss} ],
\end{equation}

where  $\mathcal{L}^\dagger$ is the Drazin Inverse of the Liouvillian $\mathcal{L}=\mathcal{L}_1+\mathcal{L}_2$:
\begin{eqnarray}
     \mathcal{L}_1\rho &=& -i\Omega \hat{J}_x \rho + \frac{1}{2} (\hat{J}_-\rho \hat{J}_+ - \hat{J}_+ \hat{J}_-\rho), 
     \\
   \mathcal{L}_2\rho &=& -i\Omega \rho \hat{J}_x + \frac{1}{2} (\hat{J}_-\rho \hat{J}_+ - \rho\hat{J}_+ \hat{J}_-).  
\end{eqnarray}
For the model under consideration, we are able to evaluate analytic expressions for the dynamical activity $\mathcal{N}$ and quantum coherence term $\mathcal{Q}$ for $N=2$ and $N=3$ cases. 
\begin{equation}
 \mathcal{N}_{N=2} = \frac{4 \gamma  \Omega ^2 \left(\gamma ^2+\Omega ^2\right)}{4 \gamma ^4+4 \gamma ^2 \Omega ^2+3 \Omega ^4},
 \quad 
 \mathcal{Q}_{N=2}=\frac{32 \left(3 \gamma ^2 \Omega ^6+\Omega ^8\right)}{16 \gamma ^7+20 \gamma ^5 \Omega ^2+16 \gamma ^3 \Omega ^4+3 \gamma  \Omega ^6}  . 
\end{equation} 
\begin{eqnarray}
   \mathcal{N}_{N=3} &=& \frac{\gamma  \Omega ^2 \left(18 \gamma ^4+12 \gamma ^2 \Omega ^2+5 \Omega ^4\right)}{18 \gamma ^6+12 \gamma ^4 \Omega ^2+5 \gamma ^2 \Omega ^4+2 \Omega ^6}
   , 
   \\
  \mathcal{Q}_{N=3} &=&  \frac{8 \Omega ^8 \left(1323 \gamma ^4+468 \gamma ^2 \Omega ^2+80 \Omega ^4\right)}{7938 \gamma ^{11}+7668 \gamma ^9 \Omega ^2+4077 \gamma ^7 \Omega ^4+1734 \gamma ^5 \Omega ^6+344 \gamma ^3 \Omega ^8+32 \gamma  \Omega ^{10}}
\end{eqnarray}
 For $N\geq 4$, quantum coherence term $\mathcal{Q}$ can be calculated numerically only due to algebraic complexity involved in finding the Drazin inverse of Liouvillian with larger dimensions. In order to illustrate our findings, We have plotted the ratio $\mathcal{N}/(\mathcal{N}+\mathcal{Q})$ in Fig. \ref{KURratio} as a function of driving strength $\Omega$ for fixed value of $\gamma$. All the curves show the same trend. As the driving strength increases, the contribution from the quantum coherence term $\mathcal{Q}$ increases and in the strong coupling limit ($\Omega>>\gamma$), contribution from dynamical activity term $\mathcal{N}$ becomes negligibly small. 

    \subsection{Precision of the clock}  
\begin{figure}
        \centering
        \includegraphics[width=0.5\linewidth]{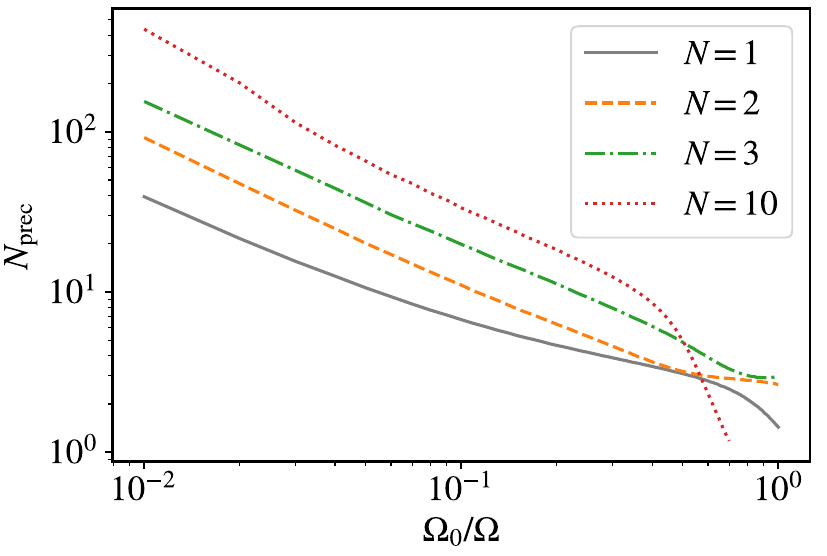}
        \caption{
        Plot of precision $N_\text{prec} = E[T]^2 / \text{Var}[T]$ versus $\Omega_0 / \Omega$. Linestyles and colors have the same meaning as Fig.~\ref{KURratio}.
        }
        \label{fig:precision}
    \end{figure}

   Since a quantum clock is subjected to  continuous measurements, the period ($T$) of the timing signal will be subjected to fluctuations induced by measurement backaction and quantum noise. Therefore, in order to characterize quantum clocks, we need to consider the average period ($E[T]$) and variance ($\text{Var}[T]$) in the period.  Precision of  a quantum clock can be defined as
    \begin{equation}
        N_\text{prec}= \frac{E[T]^2}{\text{Var}[T]}
    \end{equation}
     In Fig. \ref{fig:precision}, we have plotted the precision $N_\text{prec}=E[T]^2/\text{Var}[T]$ as a function of $\Omega_0 / \Omega$ for different values of $N$, by numerically solving Eq.~\eqref{eq:SME}. For each value of $\Omega_0 / \Omega$, we measure the mean and the variance of clock period from $10^6$ samples. This shows that precision of the clock decreases with increasing $\Omega_0 / \Omega$.

\subsection{Kinetic uncertainty relation}
\begin{figure}
        \centering
        \includegraphics[width=0.99\linewidth]{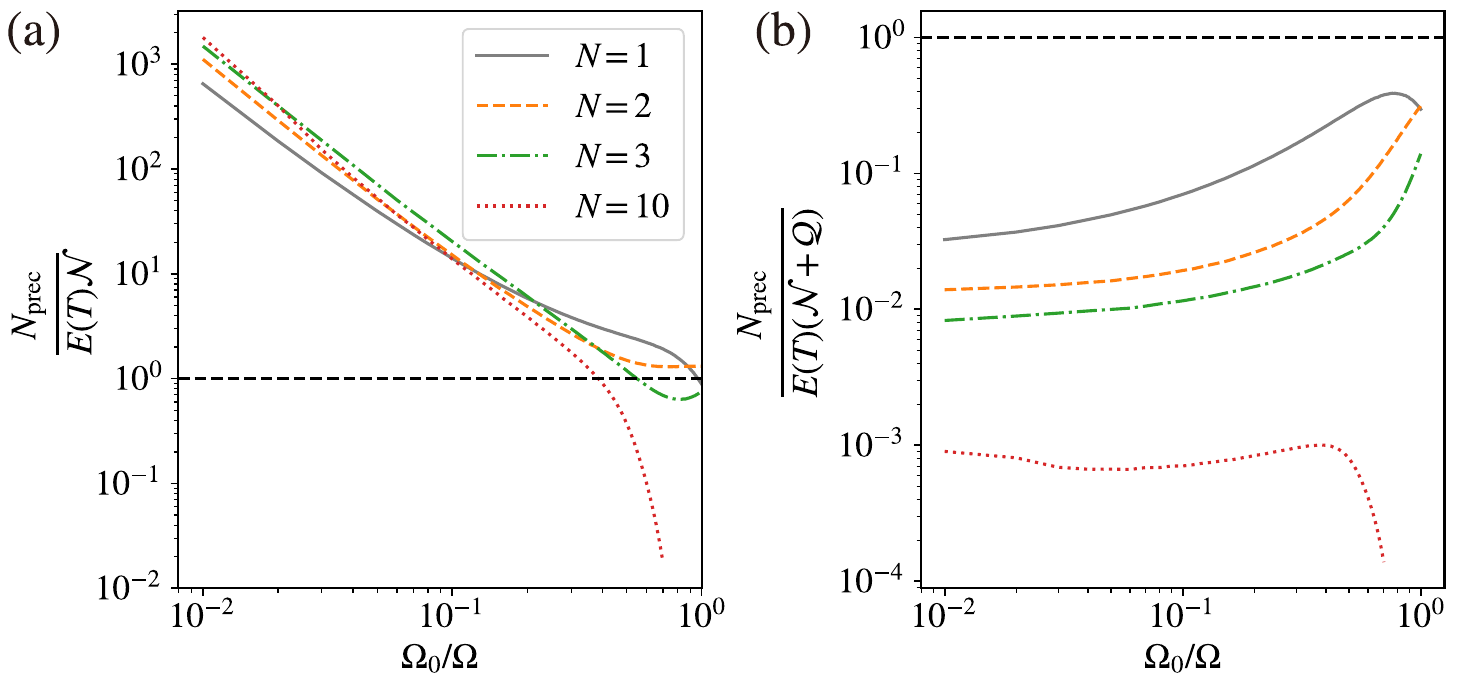}
        \caption{Illustration of (a) the classical KUR and (b) the quantum KUR. Dotted horizontal lines represent the bound, and styles and colors follow from Fig.~\ref{KURratio} and \ref{fig:precision}.}
        \label{fig:KUR_test}
    \end{figure}

 Along with the thermodynamic considerations, kinetic considerations also play fundamental role in quantum clocks operating far from equilibrium.    Quantum kinetic uncertainty relation (KUR) for the first passage time reads as:
\begin{equation} \label{eq:KUR}
       N_{\rm prec}   \leq E[T] (\mathcal{N}+\mathcal{Q}) 
\quad
        \Rightarrow   \frac{N_{\rm prec}}{E[T] (\mathcal{N}+\mathcal{Q}) }  \leq 1. 
\end{equation}
 In Fig.~\ref{fig:KUR_test}, we have numerically verified KUR (Eq.~\eqref{eq:KUR} ) by simulating Eq.~\eqref{eq:SME}  for different values of collective angular momentum $J$. The classical KUR $N_\text{prec} \le E[T] \mathcal{N}$ is severely violated for small $\Omega_0/\Omega$, but the quantum KUR (Eq.~\eqref{eq:KUR}) is always satisfied.

\section{Thermodynamics.}
Work is done on the system by an external oscillatory classical driving field and energy is dissipated through spontaneous emission into a zero temperature heat bath. In the semiclassical limit the model exhibits a Hopf bifurcation to a limit cycle as discussed. 
It forms when the total work done by a driving field over each cycle equals the energy dissipated over the cycle.

The total change in average energy due to power supplied by the driving and lost through spontaneous emission in the interaction picture is 
\begin{equation}
\frac{d\bar{E}}{dt} = \hbar\omega_a\Omega \langle J_y\rangle -\hbar\omega_a\gamma\langle J_+J_-\rangle=\hbar\omega_a\frac{d\langle \hat{J}_z(t)\rangle}{dt}
\end{equation}
The first term proportional to $\Omega$ represents the average power done by the driving field and the second term, proportional to $\gamma$, is the power lost to spontaneous emission.

\begin{figure}
    \centering
    \includegraphics[scale=0.5]{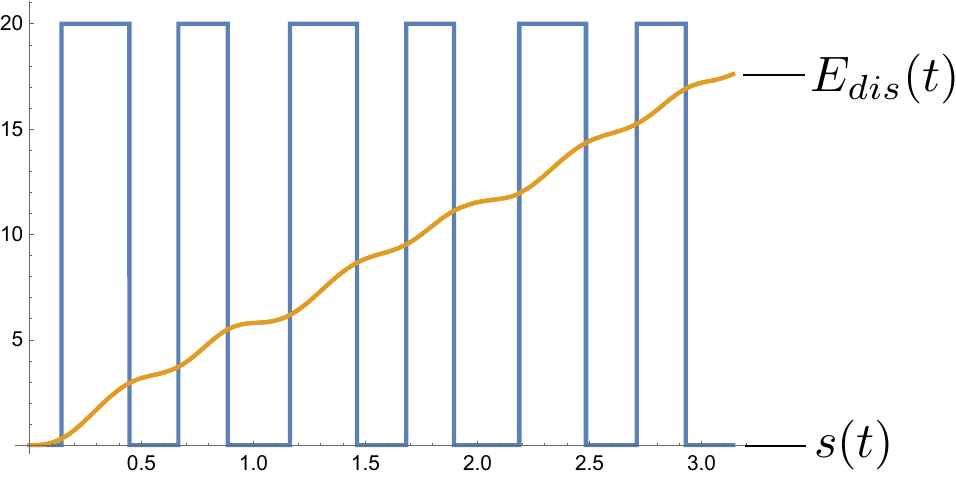}
    \caption{The energy dissipated, in units of $\hbar\omega_A$, as a function of time. The clock signal is superimposed showing the fluctuating periods of the limit cycle.  Parameters: $j=10,\ \gamma =0.1,\Omega=2\pi.$}
    \label{energy-dis}
\end{figure}

The total energy dissipated as a function of time is thus,  
\begin{equation}
    E_{dis}(t) =\hbar \omega_A \ \gamma \int_0^t \  dt \langle J_+J_-\rangle_c
\end{equation}
Using $J_+J_-= \frac{N}{2}(\frac{N}{2}+1) - J_z^2+J_z$
we see that around a cycle with a period $T$
\begin{equation}
    E_{dis}(t) =\hbar \omega_A\ \gamma\ \left [\frac{N}{2}(\frac{N}{2}+1) T - \int_{cyc} dt\  \langle J_z^2\rangle_c\right ]
\end{equation}
In the limit of large $N$, we can factorise the conditional moment in the second term. Using the semiclassical approximation for $\langle J_z^2\rangle_c$, we see that
\begin{equation}
    \int_{cyc} dt\ \langle J_z^2\rangle_c =\frac{N^2}{8}
\end{equation}
Then
\begin{equation}
    E_{dis}(t) \approx \hbar \omega_A\ \gamma\ (\frac{N^2}{4} T-\frac{N^2}{8})
\end{equation}
and we find that the average energy dissipated on a limit cycle, in the conditional evolution, is proportional to the clock signal period, $T$, a random variable. Thus the average energy dissipated on a cycle has the same statistics as the period. This is illustrated for one sample trajectory in Fig. (\ref{energy-dis}). 
From Fig. (\ref{fig:precision}) we see that the precision of the clock improves with increasing driving power and increasing $N$. The price paid for a better clock is greater power dissipation. This is the expected limit for a good clock \cite{Erker,Milburn-CP}.  

\section{Discussion and Conclusion}
An emerging perspective in thermodynamics establishes a link between precision, energy dissipation, and entropy production in classical periodic clocks\cite{Erker,Milburn-CP, Ares, quanta, nello2024}. In the classical case, this is due to thermally driven phase noise on a limit cycle such that the phase diffusion rate is inversely proportional to the size of the limit cycle. As the latter necessarily entails greater energy dissipation, we see that clock performance improves as the dissipation rate increases. In a classical limit cycle, the rate of phase diffusion is proportional to temperature, and naively it would appear that lowering the temperature would lead to perfect precision. This conclusion is not valid as, at zero temperature,  we expect quantum fluctuations to dominate.     

In this paper we have studied a dissipative non linear quantum system at zero temperature and quantified how quantum noise still leads to phase diffusion but nonetheless gets slower as the dissipation rate increases.  While the system has an exact steady state density operator above the Hopf bifurcation, a consideration of the quantum stochastic dynamics in homodyne detection of the radiated field shows that a noisy limit cycle oscillation is present and can be used to construct a  clock signal.      This validates, in a fully quantum stochastic system, the link between energy dissipation and clock precision.  

National standards laboratories worldwide have achieved remarkable levels of clock precision, raising the question of why alternative approaches are necessary. The rapid development of quantum technologies, such as quantum sensing and quantum computing, provide the motivation. The emerging risk to economic growth and security from GPS denial and spoofing is driving the search for autonomous navigation tools \cite{QED-C}. In the absence of a GPS signal,  navigation requires on board clocks, accelerometers and gyroscopes.   Quantum computing is limited by errors due to the internal clock \cite{PhysRevLett.131.160204}. Ultimately, the required precision of these devices requires quantum engineered systems that are small, portable and capable of running on low power. New design principles for quantum clocks are highly sought and the collective resonance fluorescence of this paper provides an example.

\section*{Acknowledgements.}
G J Milburn thanks Michael Kewming for useful discussions and acknowledges the support of FQXi program on {\em Information as Fuel}. Varinder Singh acknowledges the financial support through the KIAS Individual Grant No. PG096801 (V. S.) at Korea Institute for
Advanced Study.

\bibliography{fluct-quant}

\end{document}